\documentclass[12pt,preprint]{aastex}
\usepackage{graphicx}

\shorttitle{The influence of mergers on BCG mass dispersion}
\shortauthors{Jagemann}

\begin{document}

\title{The influence of galaxy mergers on the mass dispersion of brightest
	cluster galaxies}

\author{Th.\ Jagemann\altaffilmark{1}}
\affil{Astronomisches Rechen-Institut Heidelberg,\\
    M\"onchhofstra{\ss}e~12-14, D-69120 Heidelberg, Germany}
\email{thomas.jagemann@esa.int}

\altaffiltext{1}{Current address: ESA/ESTEC, Postbus 299, 2200 AG Noordwijk, The Netherlands}

\begin{abstract}

The absolute magnitude of the brightest galaxy of clusters varies remarkably 
little and is nearly independent of all other physical properties of the 
cluster as, e.g., its spatial extension or its richness. The question arises 
whether the observed small scatter is compatible with the assumption of 
dynamical evolution of the cluster. This is investigated with the help of 
statistical analysis of the results of cluster simulation. The underlying 
interaction process is merging (and also destruction) of smaller galaxies
forming the giant galaxy. The cluster itself is supposed to be in virial 
equilibrium.

We find that the evolutionary importance of merger processes grows with 
decreasing scale. Rich clusters as well as their brightest members evolve 
merely slowly whereas compact groups as well as their brightest members evolve 
more rapidly and more violently. We also find that the number of merger 
processes leading to the growth of the brightest cluster galaxy (BCG) is small 
enough to keep the BCG mass dispersion below the measured value. Our simulation 
substantiate that just the combination of the initial distribution function and 
the following merging to form the BCG can explain the remarkably small variance 
of mean BCG masses between clusters of different size and different number of 
galaxies. 
 
\end{abstract}


\section{Introduction}

The mean absolute magnitude of brightest cluster galaxies is $\langle 
M_\mathrm{VI}\rangle =-22.7 \:\mathrm{mag}$ with a dispersion of $\sigma 
(M_\mathrm{VI}) = 0.4 \:\mathrm {mag}$ (\cite{hoesseletal}). The 
distribution of BCG absolute magnitudes is well represented by a Gaussian 
distribution (e.g. \cite{postman}), and the relative variation
of luminosities yields
\begin{equation}\label{md2}
\frac{\sigma(L)}{L} = 0.5 .
\end{equation}
\cite{batcheldor} show that the luminosity dispersion in the near infrared is even less.
Due to this small intrinsic dispersion BCGs are ideal tools for cosmological 
distance determination (\cite{sandage}; \cite{postman}). Furthermore their theoretical investigation offers the 
possibility to understand the properties and evolution of the cluster itself and its
members.

\cite{sandage} points out that there is a variety of possible 
accounts for the observed scatter in BCG apparent magnitudes, but all 
contributions but intrinsic dispersion (i.e. scatter in absolute magnitude) 
seem to remain small, in particular is there no systematic trend with redshift.
At least the small total dispersion shows that any single dispersion contribution remains
small. 

There are many kinds of suggestions trying to explain the observed small scatter
in magnitudes of brightest cluster galaxies. One possibility is the existence of
a common luminosity function all galaxies of a cluster (including the brightest 
one) are drawn from (\cite{schechter&peebles}), either with
or without further development. The alternative consists of a unique creation 
process, a ``standard mold'', or a not yet discovered physical effect that 
delimits the luminosities of galaxies and cuts off the bright end of the 
luminosity function. This hypothesis is underpinned by the astonishing fact that
brightest galaxies of rich clusters are just as luminous as the brightest 
observed field galaxies (\cite{humason}). 

How special are BCGs? It has been argued that surface brightness profiles of BCGs are well fit by the same S\'ersic law that describes less-luminous spheroids and obey the same relations between fitting parameters that characterise E/S0 galaxies generally (\cite{batcheldor}). On the other hand, \cite{linden} show that BCGs contain a larger fraction of dark matter as compared to non-BCGs of the same stellar mass, influenced by the cluster environment.

X-ray observations of the intracluster medium (e.g., \cite{peres}), in particular of the cluster core, has given rise to the assumption of cooling flows depositing up to 100 M$_\odot$ yr$^{-1}$, building up the BCG over many Gyrs. However, \cite{motl} support the alternative that cores of cool gas are being built from the accretion of discrete stable subclusters. At the same time they realise that the common presence of substructure in galaxy clusters argues for accretion or merger events that occured sufficiently recently not to be erased by relaxation of the cluster ( \cite{bird}). 

Galactic cannibalism, i.e.~the accretion of the existing galaxy population through dynamical friction and tidal stripping predominantly in the evolved cluster, has been proposed by several authors (e.g., \cite{ostriker}). Further investigations by Merritt(\cite{merritt83}, \cite{merritt84}, \cite{merritt85}) and others (e.g., \cite{lauer}, \cite{dubinski}) showed, however, that galaxies are moving too fast and are significantly tidally truncated thereby suppressing dynamical friction. They concluded that the dynamical friction timescales are generally too long such that galaxies could therefore only accrete a small fraction of their total luminosity within a Hubble time. 

Galactic mergers remain as the most likely dynamical effect that dominates BCG evolution. But given a vivid merger history, it is problematical how a uniform magnitude of the brightest cluster
galaxy can get over frequent encounters while being exposed to mergers with 
other galaxies, or deletion. Tightly related to this puzzle, the question comes up whether 
the common luminosity function of all up to date analysed clusters is a direct consequence of the initial mass 
fluctuations of the early universe -- or otherwise, if the mass spectrum at that
time was an entirely other one, and evolution processes provide for the 
universality of the luminosity function.

However, the scenario seems less agitated. \cite{hoessel} estimates, 
correlating spatial extension and luminosity of BCGs, that mergers of galaxies 
of comparable size lead to the formation of the first-ranked member. As he 
divides the collapse time of a cluster by the time that pass between two close 
encounters of two relatively big galaxies, he obtains about four merger events 
forming the BCG. This result coincides with the number of observed cores in 
cD-galaxies. In the same way, the luminosity function changed only little, 
because merger processes operate effectively only on large massive galaxies 
(\cite{trevese}). 

Numerical simulations of large groups of 50-100 spherical galaxies in an equilibrium cluster have been done by \cite{funato}, \cite{bode}, \cite{garijo}. These simulations find a runaway merging process occuring near the cluster center forming a BCG. Recent analyses of the Millenium Simulation (\cite{springel05}) by \cite{lucia} indicate that, on the one hand, the most massive elliptical galaxies have the oldest and most metal rich stellar populations. On the other hand, massive ellipticals are predicted to be assembled later than their lower mass counterparts, and they have a larger effective number of progenitor systems (reaching up to $N_\mathrm{eff} \approx 5$).    

In order to investigate the influence of galaxy mergers on the BCG mass dispersion systematically and with statistical significance, we set up ensembles of Monte Carlo simulations of cluster evolution. A set of parameters that describe the initial conditions of the cluster is defined, then the cluster evolution is simulated and the evolved cluster is characterized. After that, the simulation is repeated from the beginning with the same initial parameter set, evolving to a different cluster due to the somewhat stochastic nature of galaxy encounters. The number of simulations of cluster evolution for a fixed set of initial conditions defines the ensemble size, which in turn determines the level of confidence with which parameters of the evolved cluster can be specified. Then one of the initial cluster parameters is varied leaving all other parameters fixed, and an ensemble of cluster simulations is run in the same way as described above. This parameter variation is repeated with a certain step size within an interval given by observations. The interval is supposed to cover all occurences in nature (e.g., the cluster richness) and/or uncertainties as the true value could not yet be fixed by observations (e.g., the fraction of cluster mass that is still bound to individual galaxies, \cite{springel01}). Having gone through all steps of the variation of this particular cluster parameter, the next parameter is varied leaving all others fixed. In this way, the full parameter space is covered, while all parameters are varied independently of each other. Possible correlations of initial cluster parameters (e.g., size and richness) can easily be applied afterwards by restricting the parameter space to a subspace.
The required computing power for such a large number of repeated simulations of cluster evolution over a Hubble time, covering a broad parameter space, is still beyond the capacities of present-day detailed N-body simulations. That is why a Monte Carlo code has been developed to investigate the small BCG mass dispersion, surpassing a purely analytical description (e.g., \cite{merritt83}) with the power of simulating statistically significant ensemble sizes. 

The structure of this paper is as follows: In the next section the simulation method is described. Section 3 details the parameter set and the mass dispersion. Section 4 presents the results of three simulation sets with exemplarily representative parameter sets. A discussion of the results is given in Section 5. Finally, Section 6 briefs the conclusions of this work.

\section{The model}\label{model}

Numerical simulations require the masses of the galaxies. On the presumption of
a given mass-to-light ratio $\Upsilon$, Eq.\,(\ref{md2}) is directly 
translatable into the BCG mass dispersion. Though masses resulting from 
dynamical simulations are still dependent on $\Upsilon$ once compared with the
observed luminosities, the relative variation does not depend on $\Upsilon$.

The following subsections briefly describe the model assumptions that enter the
simulation computations. For further details see \cite{jagemann}.

\subsection{The cluster}

The cluster simulation starts with $N$ galaxies homogeneously distributed in a sphere of radius $R$ and total
mass $M$. $R$ and $M$ are kept constant 
in time in the course of development; in other words, we assume that the cluster
is isolated and decoupled from the Hubble flow. At the beginning the number of 
galaxies in the cluster equals $N=N_0$. $N$ diminishes in the course of time due
to mergers and destruction of galaxies. The mass of the destroyed galaxies is 
counted to the intergalactic medium, so that the total mass of the cluster 
remains constant.
 
The galaxy masses are initially drawn from the exponential distribution function
discussed in Sect.~\ref{massdispersion}. The model starts from the assumption 
that the total mass of the galaxy cluster is proportional to the total mass of 
all galaxies:

\begin{equation}
M_\mathrm{tot} = N_0 \, M_0 \cdot k 
\end{equation}
Here $N_0$ is the initial galaxy number and $M_0$ is the mean galaxy mass 
(Sect.~\ref{massdispersion}). Then the constant $k$ is the ratio of cluster mass
to galaxy mass, i.e. the ratio of mass-to-light ratio of the cluster (in the
following denoted by $\Upsilon_\mathrm{cluster}$) and mass-to-light ratio of the
galaxies:

\begin{equation}
k=\frac {\Upsilon_{\mathrm cluster}}{\Upsilon_{\mathrm galaxies}}
\end{equation}

The total mass of a (rich) cluster is of the order of $10^{15} M_\odot$. 
Therefore the mass-to-light ratio of a cluster is within the range $[200;500]\,
\Upsilon_\odot$, and so the discrepancy between the mass-to-light ratios of
present clusters as a whole and their individual members is at least $k=5$ and 
probably more. An interesting and crucial question concerning the interaction 
between galaxies of a cluster is how much of the cluster dark matter is really 
gravitationally bound within galaxies (\cite{merritt85}), i.e. the value of $k$. Two possible 
extreme cases are imaginable: either the dark matter hides within individual 
halos of single galaxies, or these halos are smeared over the whole 
intergalactic background due to tidal processes or similar effects. As virial 
velocities of the galaxies are determined by the total cluster mass, whereas
individual galaxy masses enter decisively those equations that determine the 
interaction between two encountering galaxies, the trend resulting from the 
different extreme cases can be anticipated: 

If dark matter exists in individual halos, the merger cross section is much 
larger than in a cluster containing significantly intergalactic matter, so the 
model develops more rapidly and more violently. This work considers these two 
cases. 

\subsection{The galaxies}\label{galaxies}

In this paper individual galaxies are described by two parameters: their radius 
$r_i$ and their mass $M_i$ ($i = 1\ldots N$). Mass and radius are changing
in the course of cluster evolution; at the beginning, the masses of cluster 
galaxies are characterized by a distribution function with characteristic mass 
$M_0$ and an associated distribution function in galaxy radii with 
characteristic radius $r_0$. 

The evaluation of galaxy encounters is done under the assumption that galaxies 
are represented by Plummer spheres:

\begin{equation}
\rho (r) = \frac{3}{4\pi} \frac{M a^2}{(r^2+a^2)^{5/2}}
\end{equation}
Here $a$ is the Plummer radius, $M$ is the total mass of the galaxy and $G$ is 
the gravitational constant. Later, for the energy transfer in the tidal 
approximation, the root mean square (RMS) radius of the extended perturbed mass 
distribution is needed. So we further assume a cut-off radius $r_\mathrm{RMS} 
\approx 4.3\, a$. In the following, we call this proportional factor $\beta := 
r_\mathrm{RMS}/a$. The galaxy radii $r_i = r_\mathrm{RMS}$ are related to the
drawn masses via the Faber-Jackson-law. Calculating interaction of cluster
galaxies, the significant parameter is not the radius but rather the binding 
energy which computes for a Plummer model to:

\begin{equation}
E=-\frac{3\pi}{64}\frac{GM^2}{a}.
\end{equation}

\subsection{Encounters}

Encounters of galaxies let the model evolve. In this work, encounters are 
described by the two parameters $v$ and $B$. $v$ is the relative velocity,
and the maximum impact parameter $B$ defines the maximum separation between
two passing galaxies that is evaluated to be an encounter. Fixing the total cluster mass $M_\mathrm{tot}$ and characteristic radius $R$ to be independent 
model parameters, the mean square of the spatial velocity $v_3^2$ is set by the 
virial theorem that is assumed to be applicable to the cluster:

\begin{equation}\label{glvII}
v_3^2=f^2\frac{G\, M_0\,k\,N}{R}
\end{equation}

This procedure is in contradistinction to the procedure of \cite{krivitsky}. They find a kind of ``explosive'' merging 
without the additional constraint of virial equilibrium.

The radius $R$ of the sphere is assumed to be proportional to the 
gravitational radius with proportionality factor $f^2$. In addition, we assume the spatial virial velocity $v_3$ 
(that also determines the rate of encounters) equals $\sqrt{\langle v^{-2}
\rangle}$ that has to be put into the formula of energy transfer.

The impact parameter $b$ of the actual encounter is chosen within the limits
of $[0;B]$ such that encounters are uniformly distributed upon the maximum 
encounter cross-section $\pi B^2$. 

Using the impulse approximation, in this model $B$ is defined the way that the 
error due to neglecting encounters with impact parameters greater than $B$ is 
less than one per cent for a galaxy of standard size $r_0$ ($r_0$ corresponds 
to $M_0$ via the Faber-Jackson-law).

The mean time between two encounters is given by
\begin{equation}\label{begegnungen}
\Delta t = \frac{8R^3}{3vB^2}\frac{1}{N(N-1)}
\end{equation}
and diminishes with decreasing galaxy number $N$ during the cluster 
development. The total development time is set to $T=10^{10}\mathrm{yrs}$.

\subsection{Development}

Gravitational interaction is the origin of cluster development. We compare the
kinetic energy from orbital motion of two encountering galaxies with their 
internal energies. The binding energy of an individual galaxy increases at the 
expense of the orbital energy, i.e. at the expense of the total kinetic energy 
of their relative motion. As long as the total energy transfer $\Delta E_i + 
\Delta E_j$ is less than the kinetic energy of the relative motion, the pair of 
galaxies is assumed to stay unbound. The energy input leads to an increase in 
binding energy of a galaxy which therefore expands. If energy input exceeds 
the binding energy of the concerning galaxy, this leads to its destruction.

But if the total energy transfer is higher than their kinetic energy,

\begin{equation}
\Delta E_i + \Delta E_j > E_\mathrm{kin},
\end{equation}
the encounter ends up in a bound system, and it is assumed that the galaxies 
merge. If the encounter leads to a merger event, both participating galaxies are
replaced by a single one (the remnant), and its mass and radius has to be 
recalculated. The remnant mass is simply the sum of both galaxy masses, and the 
radius is computed from energy conservation:

\begin{equation}
E_\mathrm{R} = E_i+E_j+E_\mathrm{kin}.
\end{equation}
It is possible that the remnant binding energy becomes $E_\mathrm{R} > 0$, if

\begin{equation}
E_\mathrm{kin} > -(E_i+E_j).
\end{equation}
In this case no remnant is created and both galaxies break up.

To calculate energy exchange, the energy transfer is expressed using the impulse
and tidal approximation (\cite{spitzer}) combined with a smooth 
interpolation of the two limiting cases of $b=0$ and $b\ge \hat{b}$ ($\hat{b} =
5 \max (r_{\mathrm{h},i}, \: r_{\mathrm{h},j})$, $r_\mathrm{h}$ is the median 
radius of the respective perturbed system). 

The energy transfer of a particle of mass $M_j$ passing an extended particle 
system at distance $b$ with velocity $v$ according to impulse and tidal 
approximation is

\begin{equation}\label{spitzer1}
\Delta E_i = \frac{4G^2M^2_jM_ir^2_i}{3v^2b^4},
\end{equation}
where $r_i$ is the RMS-radius of the perturbed system $i$ (
\cite{spitzer}).

\begin{figure} 
\plotone{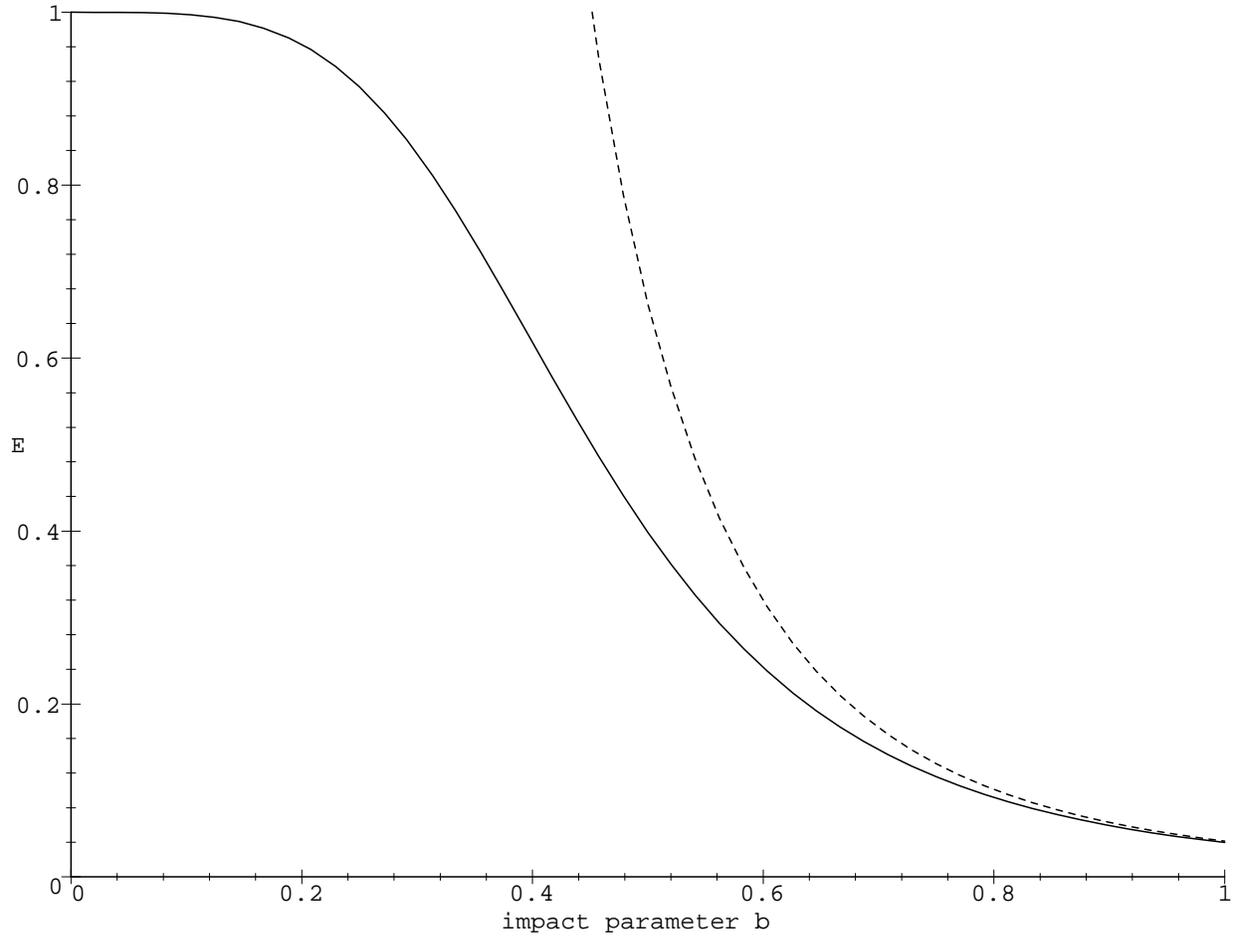}
\caption{Energy transfer with interpolation formula Eq.~\ref{interpol}. 
	Abscissa: $b/\hat{b}$, ordinate: $\Delta E/\Delta E_\mathrm{max}$; 
	dashed line: impulse approximation, solid line: interpolation.}
\label{interpolgraf}
\end{figure}

In the case of central penetration ($b=0$) the energy
transfer between two identical Plummer models is given by

\begin{equation}\label{spitzer2}
\Delta E_\mathrm{max} = \frac{G^2\beta^2m^2_jm_i}{3v^2r_i^2}
\end{equation}
where $\beta$ was introduced in Subsect.~\ref{galaxies}. The interpolation
formula is then assumed to be

\begin{equation}\label{interpol}
\frac{\Delta E}{\Delta E_\mathrm{max}} = \frac{1}{1+\frac{\beta^2}{4r_i^4}
	b^4}
\end{equation}
(Fig.~\ref{interpolgraf}).

Proceeding this way we use two terms of energy transfer in 
high speed encounters, i.e. in encounters where relative velocities are higher 
than the galaxy internal velocity dispersion. This approximation is not too 
crude (\cite{aguilar}) although galaxy relative velocities are 
about as high as their star velocities.

Other dynamical processes, especially dynamical friction between galaxies or
galaxies and the intergalactic background, tidal interaction between galaxies 
and the cluster potential as well as mass-loss due to ram-pressure are 
neglected.

\section{Mass dispersion}\label{massdispersion}

Trying to understand the BCG mass dispersion, it is necessary to know more about
the stochastic variance, inherent in the distribution function. When drawing 
$N$ galaxies from a given frequency distribution, there is already a dispersion 
in the mass of a distinct, e.g. the brightest, galaxy depending on the kind of 
distribution and the number of drawn galaxies. An analytical approximation of 
the global luminosity distribution of galaxies in today's clusters is 
Schechter's law (\cite{schechter}):

\begin{equation}
N(L) \, \mathrm{d}L = N_\ast \left( \frac{L}{L_\ast} \right)^\alpha 
	\exp \left( - \frac{L}{L_\ast} \right) \frac{\mathrm{d}L}{L_\ast} 
\end{equation}
with
\begin{eqnarray*}
\alpha & = & -5/4 \\
L_\ast & = &  1.0 \cdot 10^{10} \, h^{-2}\, L_\odot \mbox{ in the visual band}\\
N_\ast &\in& [20;115].
\end{eqnarray*}
$\alpha$ is given within an error of 20\%, but there are also votes for an
exponential distribution in compact clusters ($\alpha = 0$) differing from that
of field galaxies (\cite{geller&peebles}). In any case, solely
the number of the brightest (cD-) galaxies in the centres of galaxy clusters are
not exactly accounted for (\cite{dressler}, and references therein). This fact by itself is 
an indication of the answer to the central question: Is Schechter's law the 
result of development in kind of merger and destruction processes or does it 
reflect the mass spectrum in the early universe (\cite{lauer})? Since the BCG is in an 
exceptional position with respect to Schechter's law and does not fit the global
distribution, it is not probable that Schechter's distribution accrues from a 
different distribution (except that other than development processes would 
affect the BCG luminosity). Otherwise the distribution would include the BCG.

\begin{figure}
\plotone{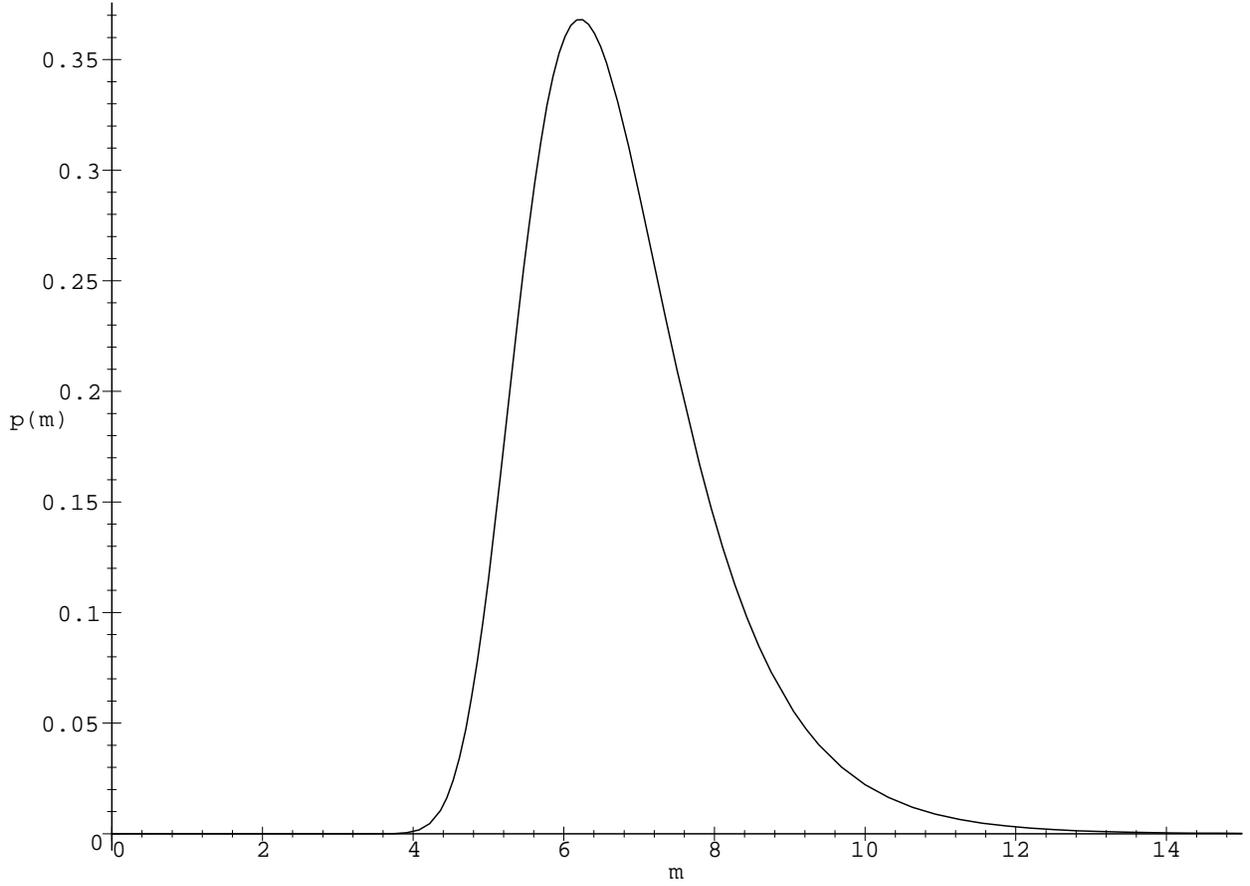}
\caption{Frequency distribution of the brightest cluster galaxies with 
	exponential mass distribution and $N=500$ and $M_0 = 1$.}
\label{f3}
\end{figure}

With the assumption of a specific mass-to-light ratio of individual galaxies,
galaxy masses can be drawn from a universal mass distribution. Because of better
analytical handling and to avoid divergences galaxy masses of the model cluster 
are drawn from an exponential distribution. In this case, the initial 
probability distribution is

\begin{equation}
	p_\mathrm{BCG}(M)\,\mathrm{d}M\equiv
	N\left( 1-{\mathrm e}^{-\frac{M}{M_0}}\right)^{N-1}
	 \frac{1}{M_0}{\mathrm e}^{-\frac{M}{M_0}}\mathrm dM,
\end{equation}
shown in Fig.~\ref{f3}.
 
In this case the expected value of the most massive initial galaxy can be 
written as an harmonic sum that depends on the mass $M_0$ characterizing the 
distribution function and on the number of drawn galaxies $N_0$:

\begin{equation}
\langle M_\mathrm{BCG}\rangle = M_0\sum_{k=1}^N\frac{1}{k}\; .
\end{equation}
The BCG mass dispersion is then computed to
\begin{equation}
\sigma_\mathrm{BCG} = M_0\sqrt{\sum_{k=1}^N\frac{1}{k^2}}\; .
\end{equation}

With $N=1000$, e.g., the relative deviation computes to $\sigma_\mathrm{BCG}
/\langle M_\mathrm{BCG}\rangle = 0.17$. Though the expected value of the BCG
masses increases with galaxy number, the relative deviation does not. Generally
the fact is remarkable, that initially BCG masses depend on the number of 
galaxies, whereas in reality such a dependence is not found, and the BCG 
relative mass dispersion is distinctly smaller for all numbers of galaxies than
the observed one. These differences may be due to development processes, 
although additional correlations among $R$ and $N$ may cause selection effects 
in $(N,R)$-space.

\begin{figure} 
\plotone{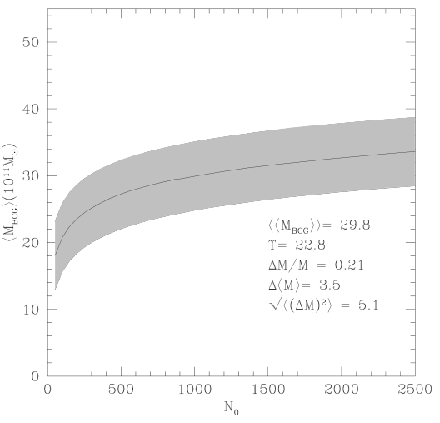}
\caption{BCG masses and their variance at the beginning of the simulation. 
	Initial parameters are: $M_0 = 4\cdot 10^{11} M_\odot, r_0=15 \,
	\mathrm{kpc}, v_0=800\, \mathrm{km/s}, R = 3 \,\mathrm{Mpc}, T=10 \,
	\mathrm{Gyr}.$}
\label{f1}
\end{figure}

Statistical statements about clusters of galaxies are derived from model 
assumptions depending on several parameters. Thus the resulting quantities to be
analysed are only estimates of the mean sampled over infinite runs. The 
accuracy, given as relative 1$\sigma$ variance, is proportional to the square 
root of the number of simulation runs. Differently set up model clusters with 
the same set of initial parameters are here called an ensemble. So since all 
our simulation results are based on ensemble sizes of 100 runs, they are 
accurate to within 10\%.

Additionally, any dispersion in absolute magnitudes (or masses, respectively) 
in a model that depends on variable parameters will generally depend on these 
parameters, too. For example, BCG masses -- as well as their dispersions -- that
are drawn from a distribution function depend on a characteristic mass $M_0$
scaling this function and depend also on the total number of galaxies $N_0$
drawn from this distribution.

\begin{figure} 
\plotone{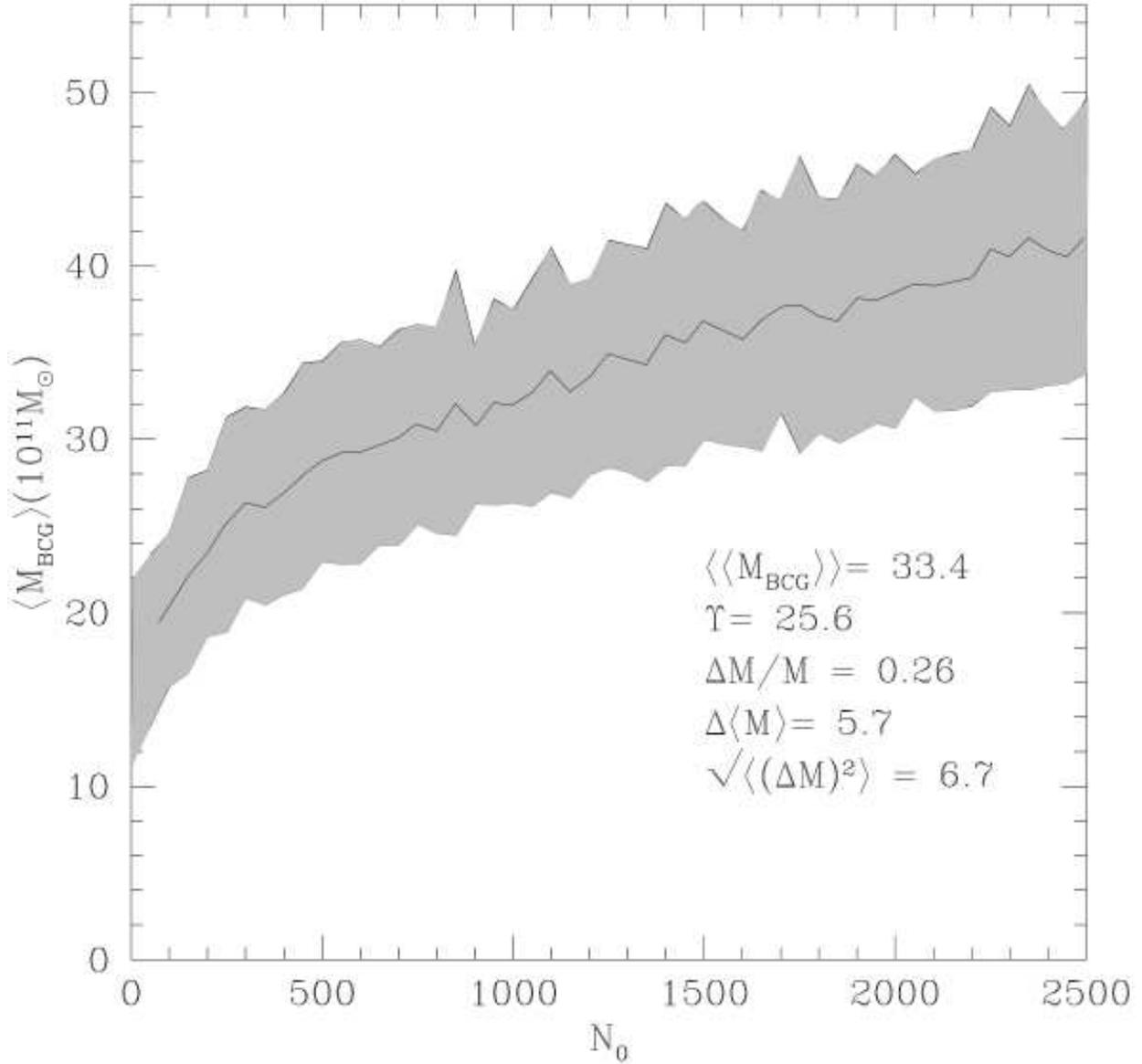}
\caption{BCG masses and their variance at the end of the simulation. The initial
	parameters are the same as in Fig.~\ref{f1}.}
\label{f2}
\end{figure}

Figures \ref{f1} and \ref{f2} illustrate the dependence of the mean of BCG 
masses on the initial number of galaxies. Likewise, standard deviations belonging to any 
value of $N$ are shown, too. Figure \ref{f1} describes the initial state, 
Fig.~\ref{f2} the evolved state. The curve in Fig.~\ref{f1} is smooth, because 
the expected value of BCG mass and dispersion are theoretically computable. In 
Fig.~\ref{f1} the stochastic 10\% noise of the mean values is obvious.

In order to calculate the total dispersion of BCG masses, let these masses be
a function of a vector $\vec{x}$ in parameter space. Furthermore, assume that a 
distribution function $p_{\vec{x}} (M)$ is given. The expected value of $M$ then 
averages the ensemble and the parameter space to

\begin{equation}
\langle \langle M\rangle\rangle\,:=\int_{\vec{x}}\langle M(\vec{x})\rangle\, 
	p(\vec{x})\,\mathrm{d}\vec{x}\, ,
\end{equation}
where $\langle M(\vec{x})\rangle$ means an ensemble average. The mass dispersion
averages the ensemble and the parameter space to 

\begin{equation}
(\Delta M)^2_\mathrm{tot} = \langle(\Delta M)^2\rangle + \langle\langle M\rangle
	^2\rangle-\langle\langle M\rangle\rangle^2,
\end{equation}
where the inner brackets always denote an ensemble average. Thus, the total 
dispersion is the expected value of the individual dispersion in $\vec{x}$
plus the scatter of the expected values of $M$ in $\vec{x}$:

\begin{equation}
(\Delta M)^2_\mathrm {tot} = \langle(\Delta M)^2\rangle + 
	(\Delta\langle M\rangle)^2.
\end{equation}
This expression is used later on to calculate the mass dispersion of the 
simulations. At the beginning of the dynamical development, any scatter in BCG 
absolute magnitudes or masses is the sum of different kinds of model dispersion:

\begin{enumerate}
\item The statistical variation at constant model parameters within an ensemble,
	whereby one gets a convolution of two distributions: the stochastic
	variance at a constant vector of model parameters and
\item the variance caused by the distribution of clusters upon the model
	parameters.
\end{enumerate}
Hence the observed relative mass dispersion is the upper limit of the 
statistical error. 

In this context it is astonishing that there is definitely no correlation 
between $M_\mathrm{BCG}$ and richness (meaning the final galaxy number) in 
today's observed clusters (\cite{sandage}; \cite{postman}). But drawing galaxies from a distinct distribution (except
from a $\delta$-function), there is always -- within the bounds of resolution --
such a dependence, so that there inevitably has to be another process to explain
the independence of $M_\mathrm{BCG}$ on $N$. The simulations try to answer the
question whether mergers can account for such a behaviour. The observed scatter 
in absolute magnitudes is in any case an upper limit of the intrinsic BCG mass 
dispersion within any ensemble. 

The dependence of the BCG absolute magnitudes on the scaling mass $M_0$ is more 
complicated. $M_0$ is positively correlated with $M_\mathrm{BCG}$ with high 
statistical significance. In order to eliminate any photometric calibration 
error as well as a global shift of the luminosity function, \cite{trevese} compare $M_\mathrm{BCG} - M_{10}$ ($M_{10}$ is the 
magnitude of the 10th brightest galaxy) with $M_0 - M_{10}$. They find, that both 
differences are negatively correlated.

\section{Results}\label{results}

Table \ref{tabelle} lists the input parameters of the simulation.

\begin{table}
\caption[ ]{Simulation parameters}
\begin{flushleft}
\begin{tabular}{lll}
\hline\noalign{\smallskip}
Parameter & Value(s) & Remark\\
\noalign{\smallskip}
\hline\noalign{\smallskip}
Ensemble size & $n=100$ & constant \\
Total time &  $T = 10$ Gyr & constant \\
Char. mass & $M_0\in\{5;50\}\cdot 10^{11}M_\odot$ & variable (1) \\
Char. radius & $r_0 \in \{20;100\}$kpc & variable (2) \\
$\Upsilon$ ratio & $k\in\{1;10\}$ & variable (1) \\
Cluster radius & $R\in [ 0.25; 3.00]$ \`a 0.25 Mpc & variable (2) \\
Richness & $N \in [ 100; 1000]$ \`a 100 & variable (2) \\
\noalign{\smallskip}
\hline
\end{tabular}
\end{flushleft}
\label{tabelle}
\end{table}

Here the two parameters $R$ and $N$ are allowed to vary independently (2) in the
given interval while the pair $(M_0;r_0)$ remains constant. Then, the mean relative 
velocity computed using the virial theorem adjusts automatically. Afterwards a 
new pair $(M_0;r_0)$ is fixed from the variation interval, and $N$ and $R$ are 
varied again. $M_0$ and $r_0$ are not correlated. $k$ is set to 1 or 10 (the 
probable value of present clusters).

The underlying idea is that there is a fixed value of $M_0$ for all clusters
in nature within narrow limits (corresponding to the fixed $L_\ast$ in the
Schechter distribution), that the mass-to-light ratio of all clusters is
initially the same and that all galaxies of equal initial mass are of the
same size, independent of cluster membership.

On the other hand, there are clusters of different size and variable richness 
in nature that enter the measurements of mass dispersion of the brightest 
cluster galaxies and are equally put together.

At the end of the simulation the following ensemble mean data are derived:

\begin{enumerate}
\item The mean relative velocity calculated from the virial theorem,
\item the number of galaxies at the end of the simulation,
\item the total number of encounters within the maximum impact parameter,
\item the number of constructive merger processes,
\item the number of destructive merger processes,
\item the number of destroyed galaxies without forming a remnant,
\item the number of merging events that lead to the formation of the BCG,
\item the BCG mass,
\item the BCG mass ensemble dispersion.
\end{enumerate}

Except for the last two items giving the relative mass dispersion, the
data serve as investigation of the model and are not connected to observations.

\subsection{The simulation with $M_0= 5\cdot 10^{11}M_\odot$, $r_0=20
	\mathrm{kpc}$ and $ k=1$}\label{simI}

In the following, results of this parameter set are discussed in detail. 
Concerning other values of $M_0$, $r_0$ and $k$ only occuring changes in 
contrast to this set are mentioned. 

\begin{figure}
\plotone{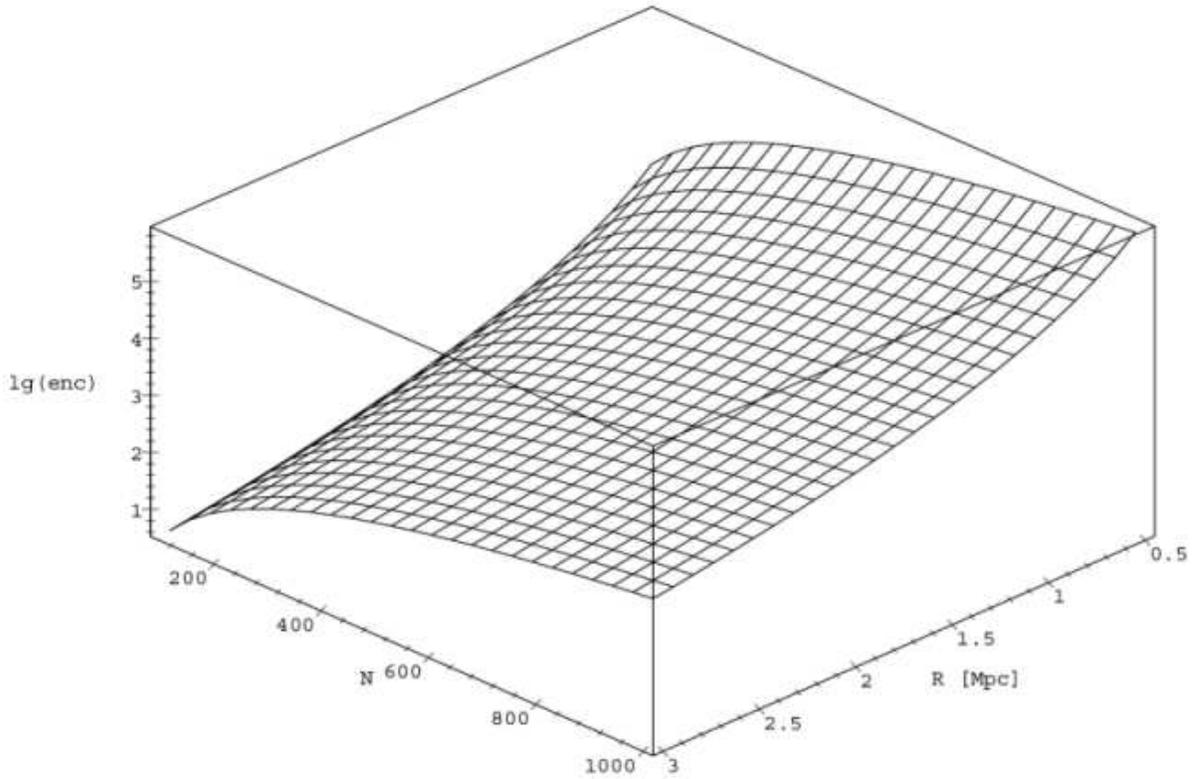}
\caption{Logarithm of the total number $p$ of encounters.}
\label{Abb9}
\end{figure}

The choice of  $M_0 = 5\cdot 10^{11}M_\odot$ means a moderate mass-to-light
ratio of about $\Upsilon_{\mathrm galaxies}=10$ of today's observed clusters,
i.e. the absence of dark coronae, and $k=1$ means the initial absence of 
intergalactic matter. 

\begin{figure}
\plotone{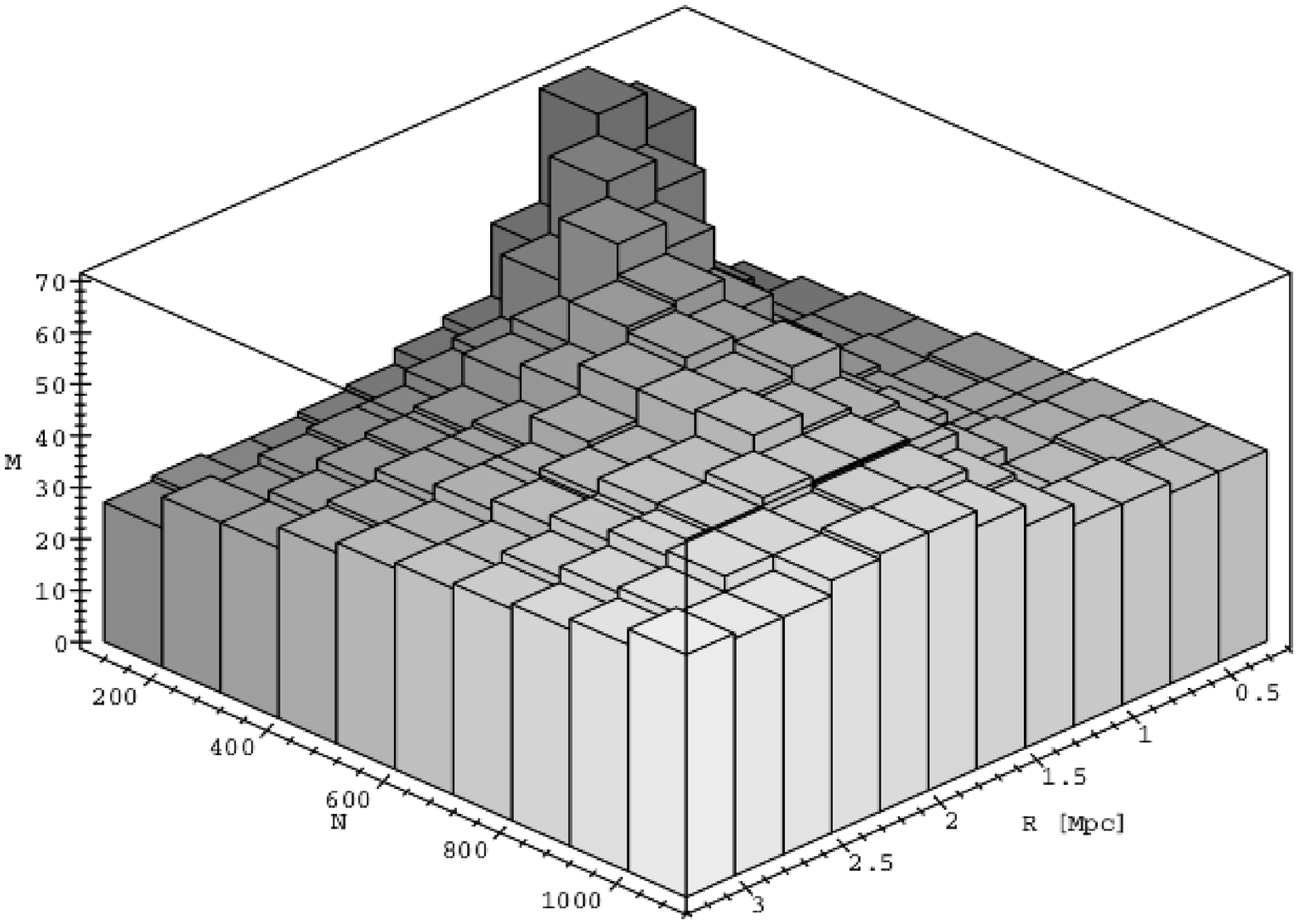}
\caption{Mean values of BCG masses after the development. $M=\langle 
	M_\mathrm{BCG} \rangle [10^{11} M_\odot]$.}
\label{Abb2}
\end{figure}

\begin{figure}
\resizebox{\hsize}{!}
{\includegraphics[angle=270]{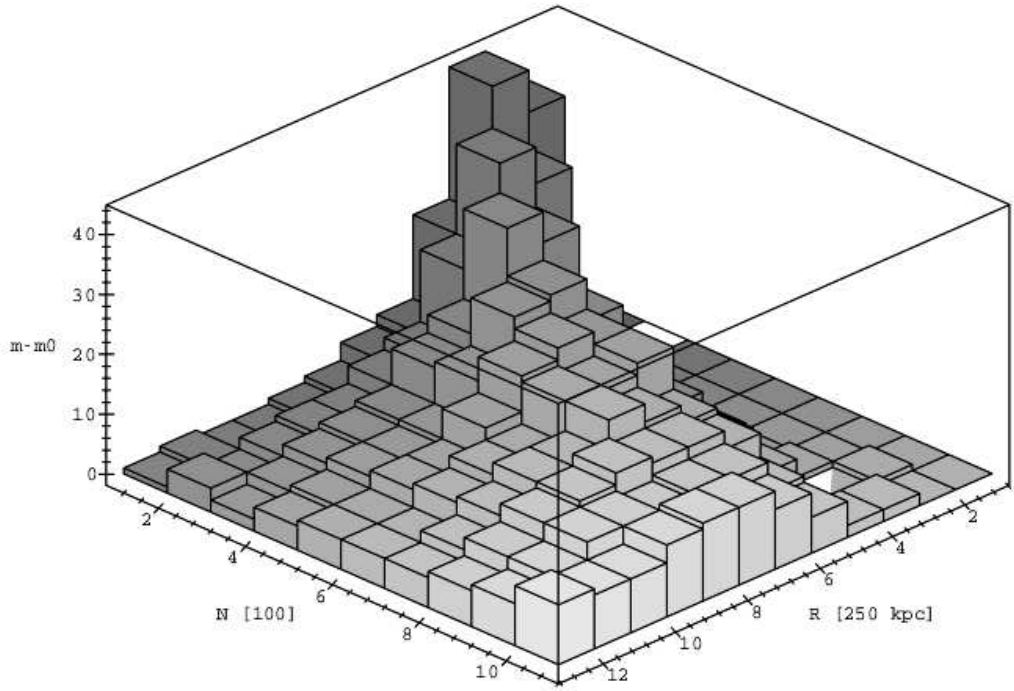}}
\caption{Mean values of BCG mass increase. $M-M0=\langle 
	M_\mathrm{BCG,afterwards}\rangle -\langle 
	M_\mathrm{BCG,previously}\rangle [10^{11} M_\odot]$.}
\label{Abb3}
\end{figure}

\begin{figure}
\plotone{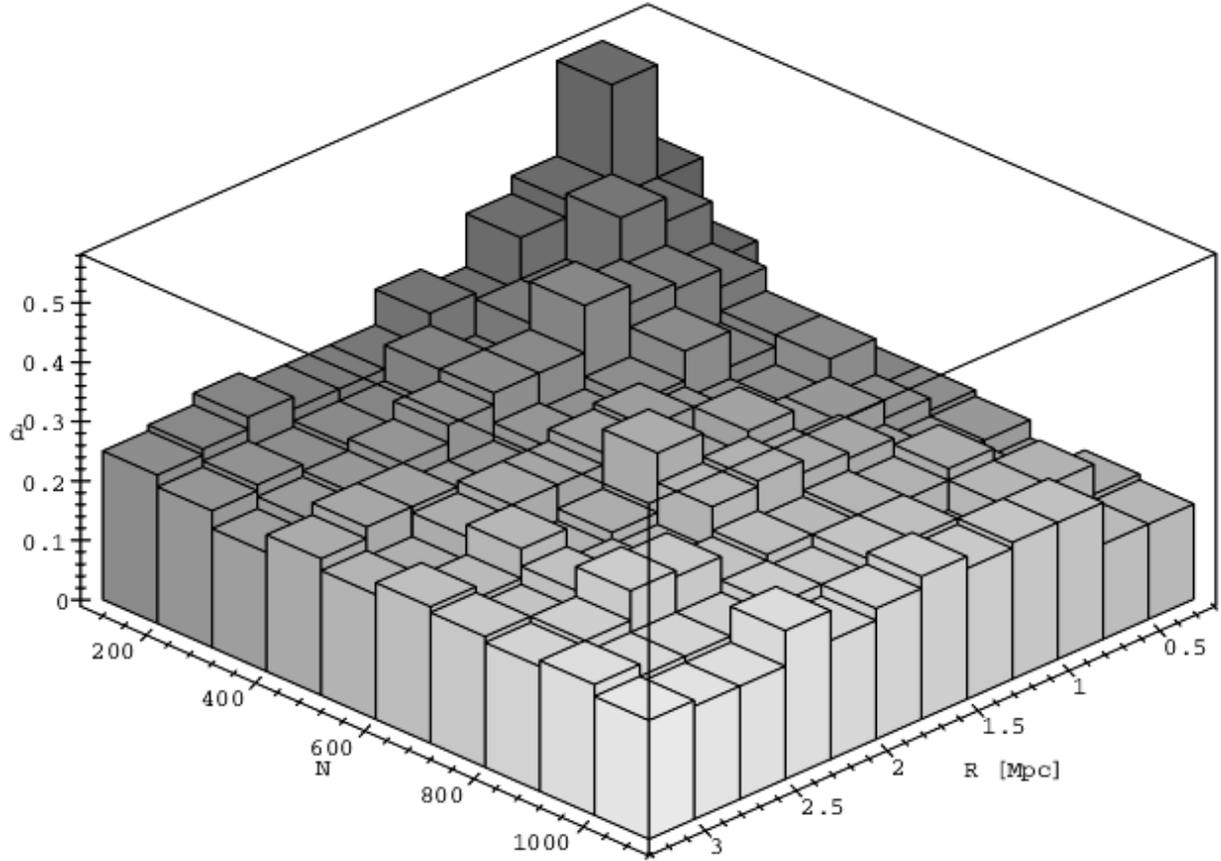}
\caption{Mean values of BCG mass dispersion after the simulation. $d=\langle 
	\sigma_\mathrm{BCG}\rangle /\langle M_\mathrm{BCG}\rangle$.}
\label{Abb4}
\end{figure}

\begin{figure}
\plotone{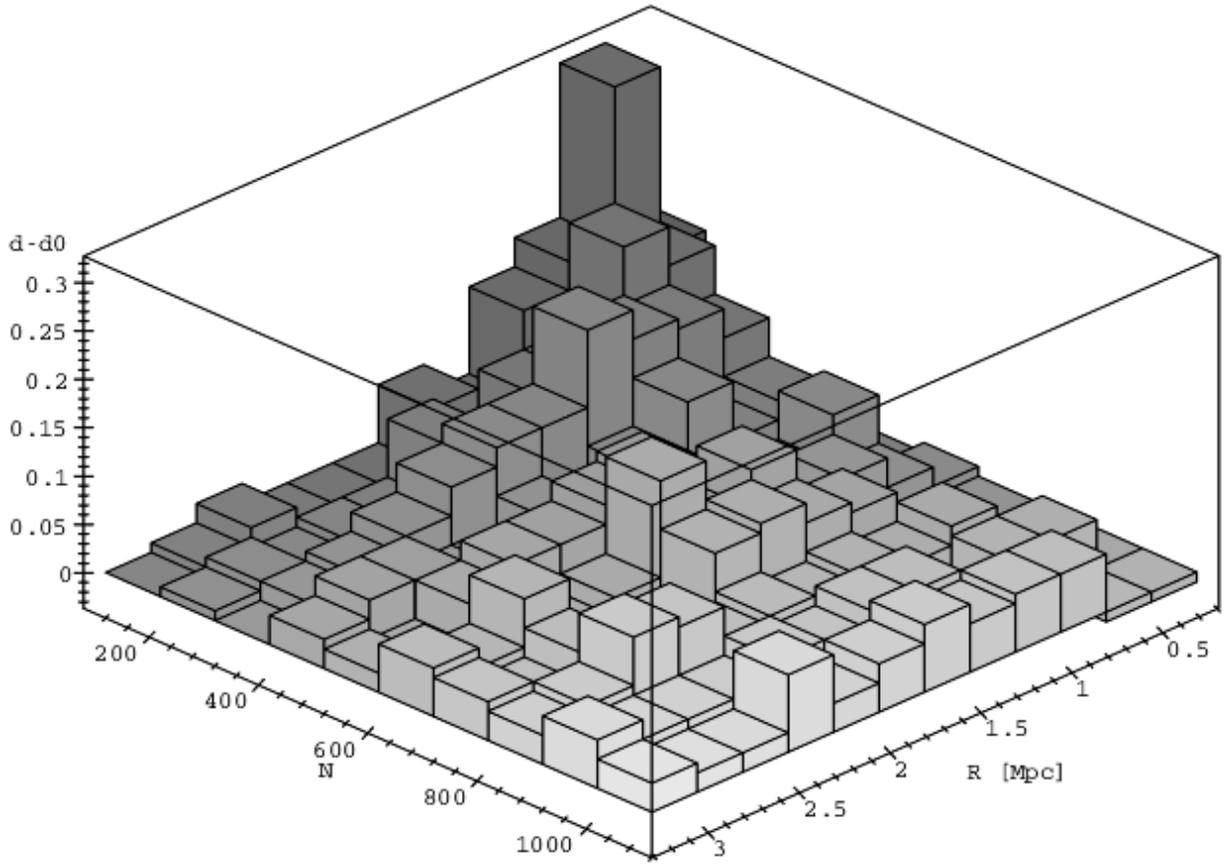}
\caption{Mean values of the increase of BCG mass dispersion. $d-d0=\langle 
	\sigma_\mathrm{BCG,afterwards}\rangle /\langle 
	M_\mathrm{BCG,afterwards}\rangle - \langle 
	\sigma_\mathrm{BCG,previously}\rangle /\langle 
	M_\mathrm{BCG,previously}\rangle$.}
\label{Abb5}
\end{figure}

\begin{figure}
\plotone{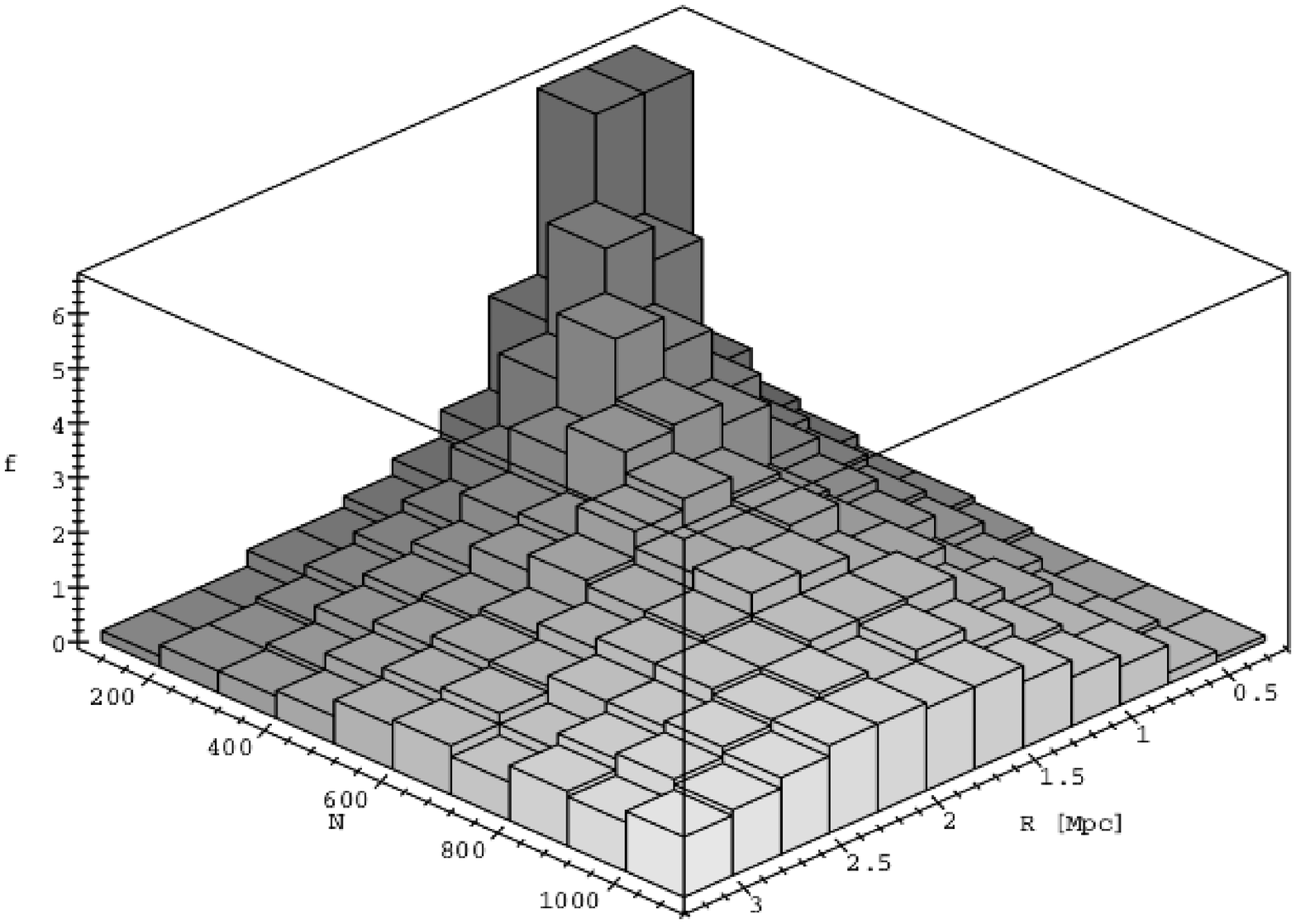}
\caption{Mean number of mergers leading to the BCG formation.}
\label{Abb6}
\end{figure}

The number of encounters of all galaxies is shown in Fig.~\ref{Abb9}. This total
number of encounters increases within the bounds of parameter variation to 
$10^6$ and corresponds essentially to Eq.~\ref{begegnungen}, substituting the
total development time $T$ for $\Delta t$. Therefore the number of encounters
is proportional to $N^{5/2}$ and to $R^{-7/2}$ and changes in the course of 
evolution only at small radii $R$ , where the galaxy number decreases due to
merger and destruction processes (cf. Fig.~\ref{Abb6}).

Figure \ref{Abb2} shows the mean BCG masses at the end of the simulation. They 
are located between about 30 and $70 \cdot 10^{11}M_\odot$ and show a clear peak
at small radii and small initial galaxy numbers.

In order to exclude the influence of the initial mass distribution in this histogram,
the pure mass increase is shown in Fig.~\ref{Abb3}. Generally the BCG is 
growing in the course of development. The increase amounts to 40 $\cdot 
10^{11} M_\odot$ at the peak which stands out even more clearly than in 
Fig.~\ref{Abb2}.

The main result of the simulation, i.e. the mean values of the BCG relative mass
dispersion, is shown in Fig.~\ref{Abb4}. The dispersion is nearly constant at 
about thirty per cent and increases at the small clusters to a maximum of 57 \%.

In order to investigate the increase in mass dispersion as in Fig.~\ref{Abb3} 
and to exclude the influence of the initial distribution, the relative mass 
distribution after the simulation is subtracted by the relative dispersion 
before the simulation in Fig.~\ref{Abb5}. As in Fig.~\ref{Abb3}, the peak 
stands out more clearly against the stochastic noise.

To answer the question how many mergers are necessary to form the BCG this 
parameter $f$ is shown in Fig~\ref{Abb6}. The maximum value lies at about six 
mergers and decreases steeply from there to a value of about one.

\begin{figure}
\plotone{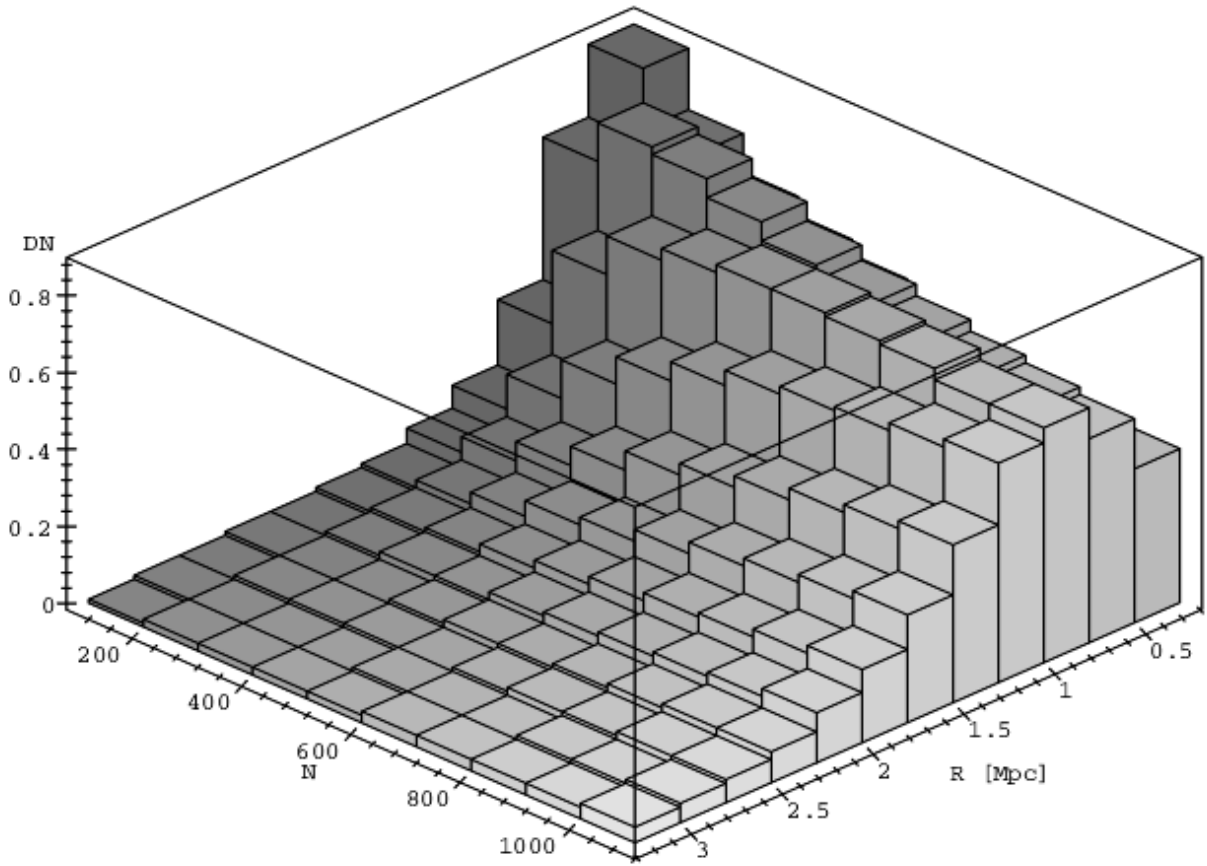}
\caption{Mean number of destroyed/merged galaxies per initial galaxy number
	after development.}
\label{Abb7}
\end{figure}

Not only merger events but also destructions of galaxies are part of cluster 
development. Figure \ref{Abb7} shows how many galaxies are left at the end of 
the simulation. Here a clear trend is recognizable: the percentage of destroyed 
or merged galaxies increases considerably with decreasing cluster radius. In 
the case of rapid evolution the number of galaxies may decrease down to twenty 
per cent of the original value.

\begin{figure}
\plotone{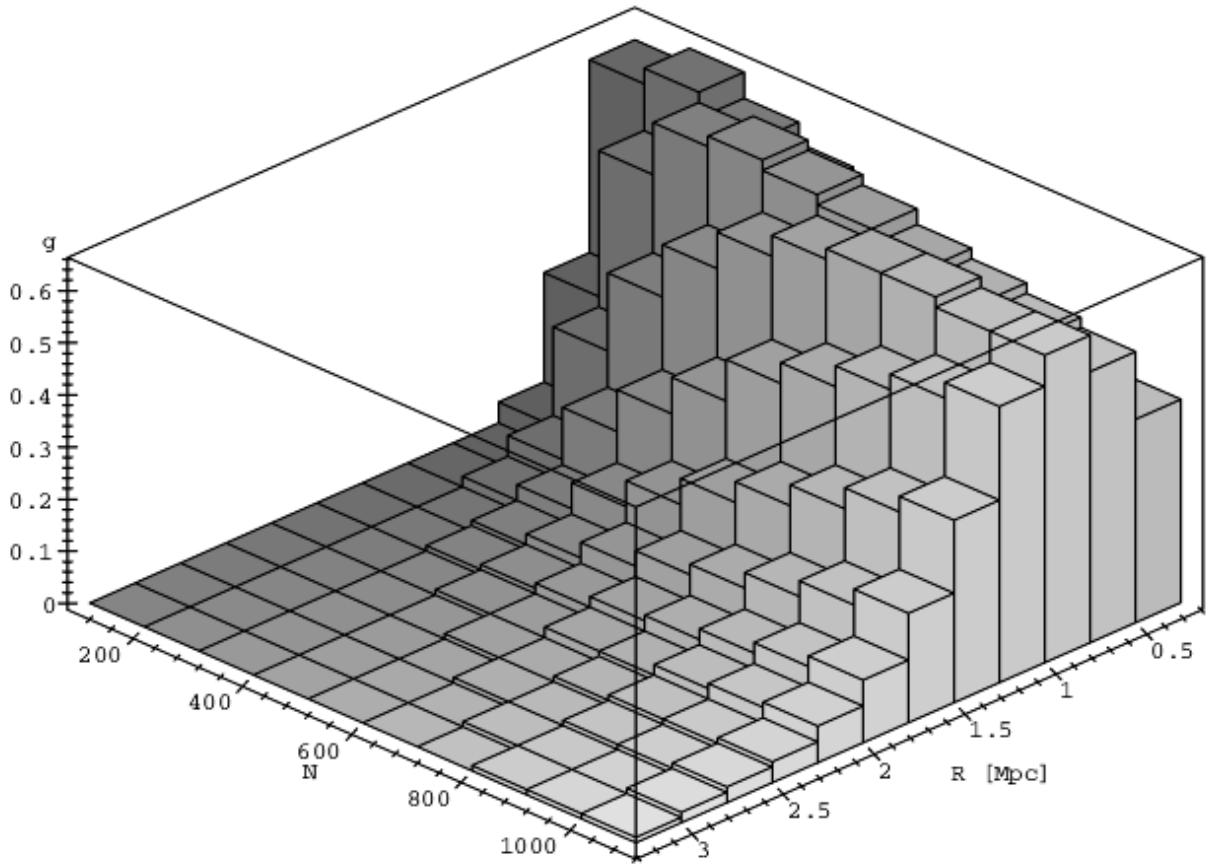}
\caption{Percentage of destroyed galaxies.}
\label{Abb10}
\end{figure}

\begin{figure}
\plotone{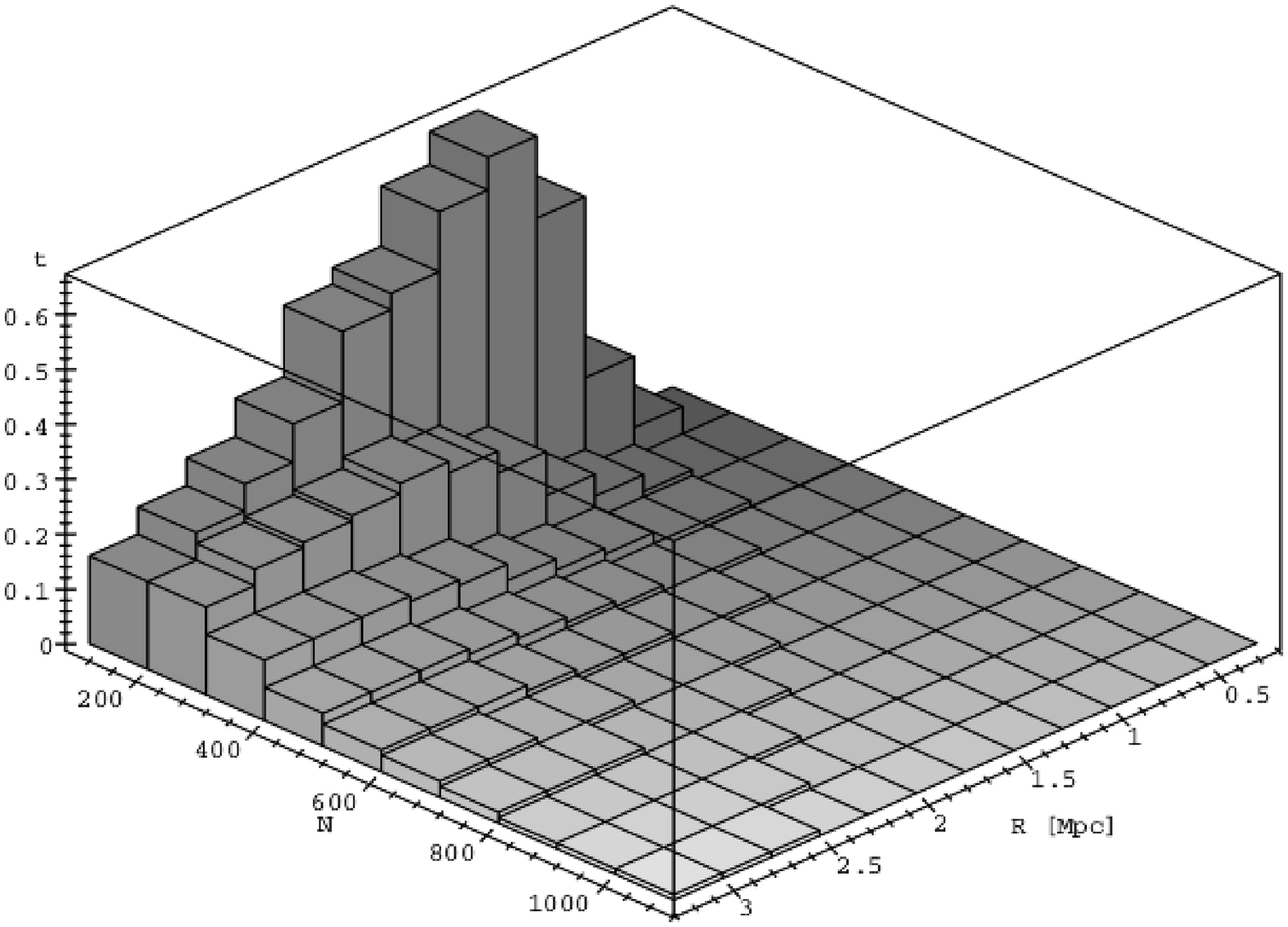}
\caption{Mean number of merger events per final galaxy number.}
\label{Abb8}
\end{figure}

The difference between Fig.~\ref{Abb7} and Fig.~\ref{Abb8} provides roughly for
the number of galaxies destroyed in encounters (Fig.~\ref{Abb10}). It turns out
that the biggest fraction of galaxies is destroyed when the energy transfer is 
less than the sum of binding energies of the 2 galaxies. The percentage of 
disrupted galaxies is always less than 65 per cent and decreases with 
increasing radius. At small radii the fraction of destroyed galaxies is roughly 
independent of the initial galaxy number, but increases with galaxy number at
increasing radii.

Finally, to address the question about the fate of the remaining galaxies, the 
number of all the merger events of all galaxies is divided by the final galaxy 
number in Fig.~\ref{Abb8}. The parameter $t$ in this diagram denotes the 
percentage of those galaxies that are involved in a merger event. In the same 
way $t$ is a measure of the frequency describing how often a galaxy has been 
involved in a merger event averaged over all cluster galaxies. In the peak 
vicinity every galaxy has experienced one merger on the average. $t$ sharply
decreases with increasing $N$ and $R$.

\subsection{The simulation with $M_0= 5\cdot 10^{11}M_\odot$, $ r_0=20
	\mathrm{kpc}$ and $ k=10$}\label{simII}

The choice of $M_0$ and $r_0$ corresponds to the choice in Sect.~\ref{simI}, $k=10$
corresponds (at $M_0= 5\cdot 10^{11}M_\odot$) roughly to the maximum value of the
present fraction of intergalactic dark matter in galaxy clusters.

The mass distribution of single galaxies corresponds exactly to the one of 
Sect.~\ref{simI}, but because of the addition of intergalactic matter the virial
velocity is increased by the factor $\sqrt{10}$ according to Eq.~\ref{glvII}. 
This is why the number of encounters is about this factor higher in comparison 
with Fig.~\ref{Abb9}. The high spatial 
velocity-dispersions a little above the diagonal of more than 2400 km/s are not 
observed in reality; therefore this part of the simulation is excluded in the 
following discussion.

The final galaxy numbers drop to values of significantly less than hundred per 
cent at small radii. 
The qualitative behaviour follows the one of Fig.~\ref{Abb7}, but the maximum 
is, however, clearlier marked at small radii and galaxy numbers, and about half 
as many galaxies less have been destroyed at a fixed pair ($R$,$N$) compared 
with Sect.~\ref{simI}. The dominant development process here is not merging but 
destruction of encountering galaxies. The mean mass distribution of BCGs after the simulation has not changed much: 
The initial mass distribution can easily be recognized. 
Deviations in comparison to the initial distribution arise in about half of the 
cases, these deviations are about  $\pm 1\cdot10^{11}M_\odot$ for each half.

Only little signs of evolution are present in the BCG mass dispersion after the simulation: The initial mass scatter is 
still present. The relative dispersion does not exceed a value of
25 per cent. The trend of the initial distribution having smaller spreads with 
increasing galaxy numbers still exists. Just as for the mean mass values there 
is an irregular change in the scatter of 1--2\% compared with the initial 
scatter.
 
\subsection{The simulation with $M_0= 50\cdot 10^{11}M_\odot$, $r_0=100
	\mathrm{kpc}$ and $k=1$}\label{simIII}

The total mass of the galaxy cluster is the same here as in Sect.~\ref{simII}, 
the galaxies however are ten times more massive. $k=1$ means the initial absence
of intergalactic matter as in Sect.~\ref{simI}. The dark matter is now hidden 
in individual halos instead of spread throughout the intergalactic medium. Therefore more extended galaxies with a characteristic radius of $r_0=100$ kpc are 
chosen at this parameter set. 

\begin{figure}
\plotone{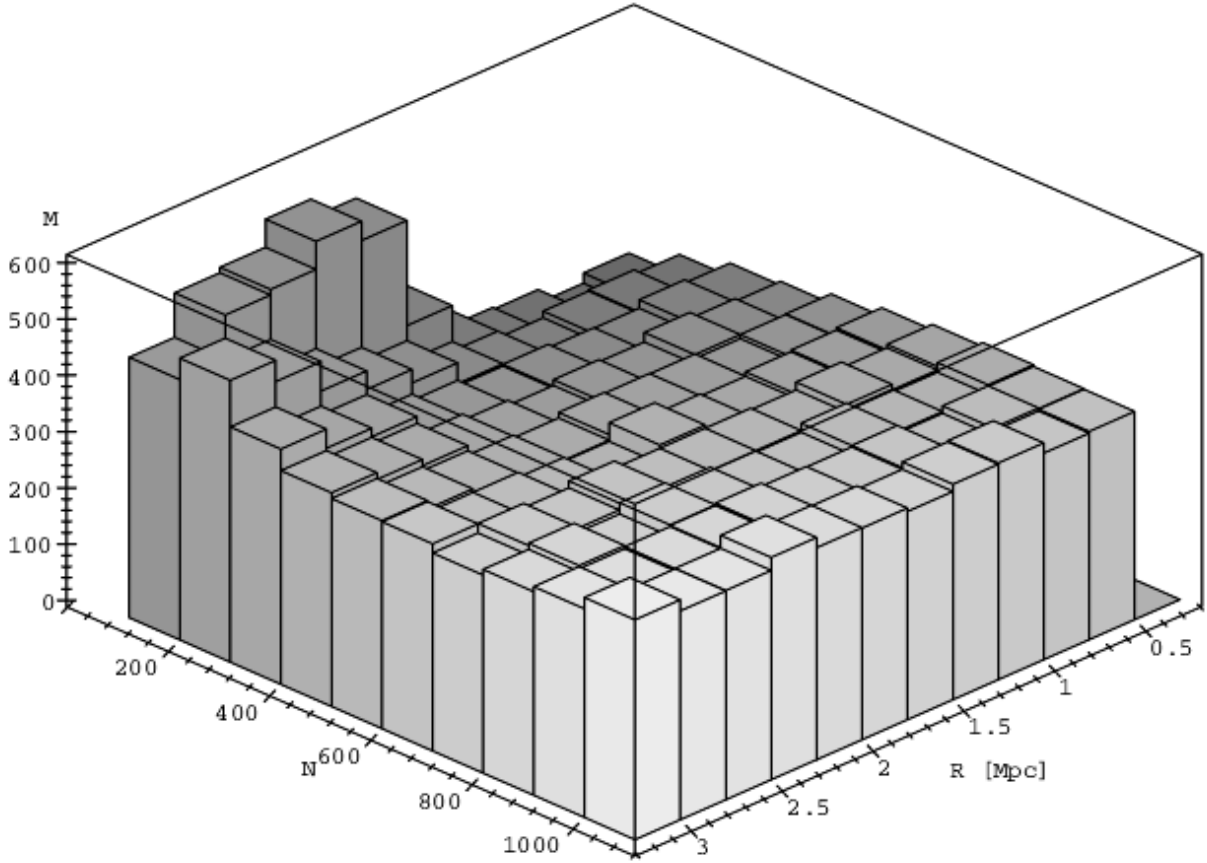}
\caption{Same as Fig.~\ref{Abb2}, but now for model \ref{simIII}.}
\label{Abb22}
\end{figure}

Figure \ref{Abb22} shows the BCG mass mean values after the development. 
In comparison with Fig.~\ref{Abb2} the shift of the maximum toward
smaller galaxy numbers but significantly greater cluster radii is striking.

\begin{figure}
\plotone{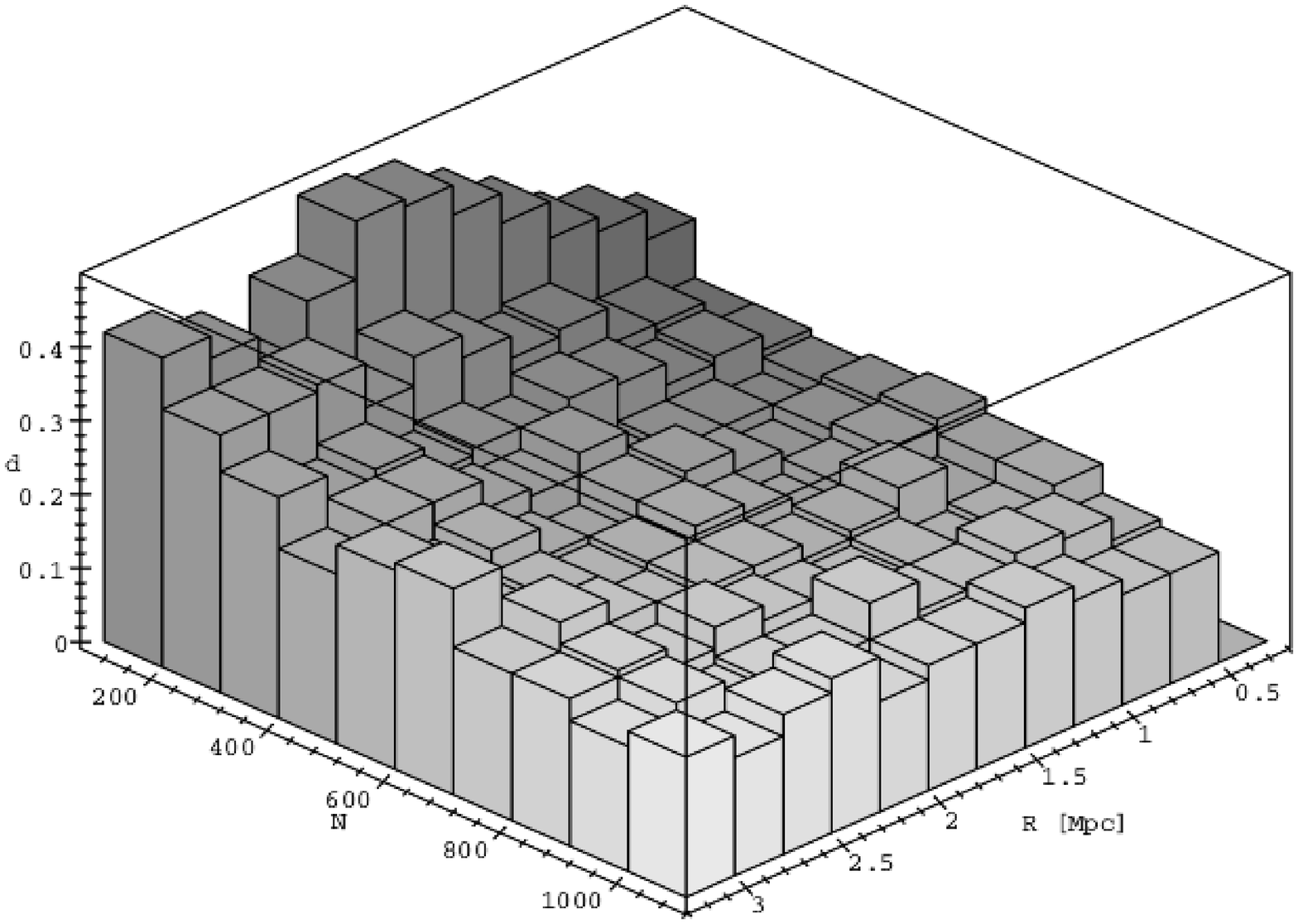}
\caption{Same as Fig.~\ref{Abb4}, now for model \ref{simIII}.}
\label{Abb24}
\end{figure}

Figure \ref{Abb24} makes it clear that the relative mass dispersion remains always
below forty per cent and even decreases significantly going to greater galaxy 
numbers, particularly at this set of parameters. 

\begin{figure}
\plotone{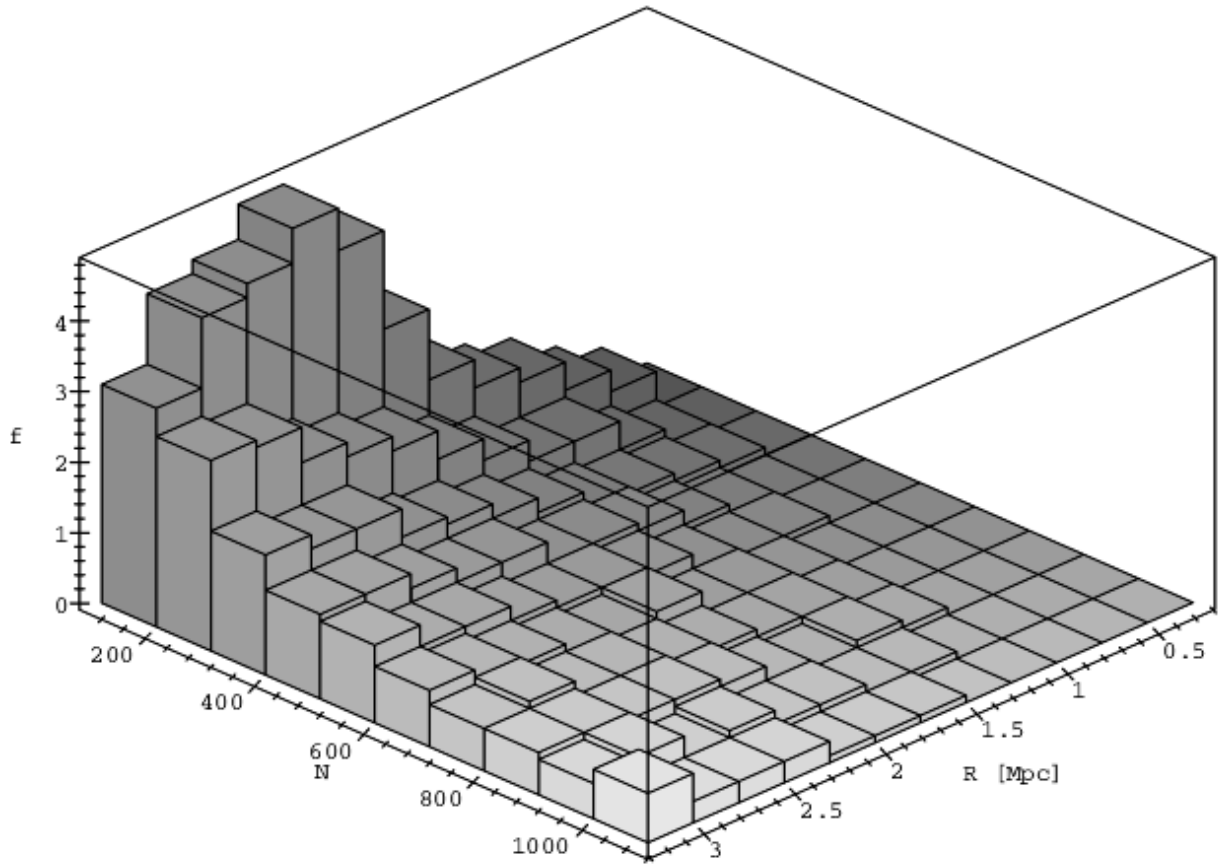}
\caption{Same as Fig.\ref{Abb6}, now for model \ref{simIII}.}
\label{Abb26}
\end{figure}

Fig.~\ref{Abb26} shows that the number of mergers leading to the formation of 
the BCG is non-zero for a wider range of parameters $N$ and $R$. The maximum 
mean number of mergers is four as in Sect.~\ref{simI}.

\begin{figure}
\plotone{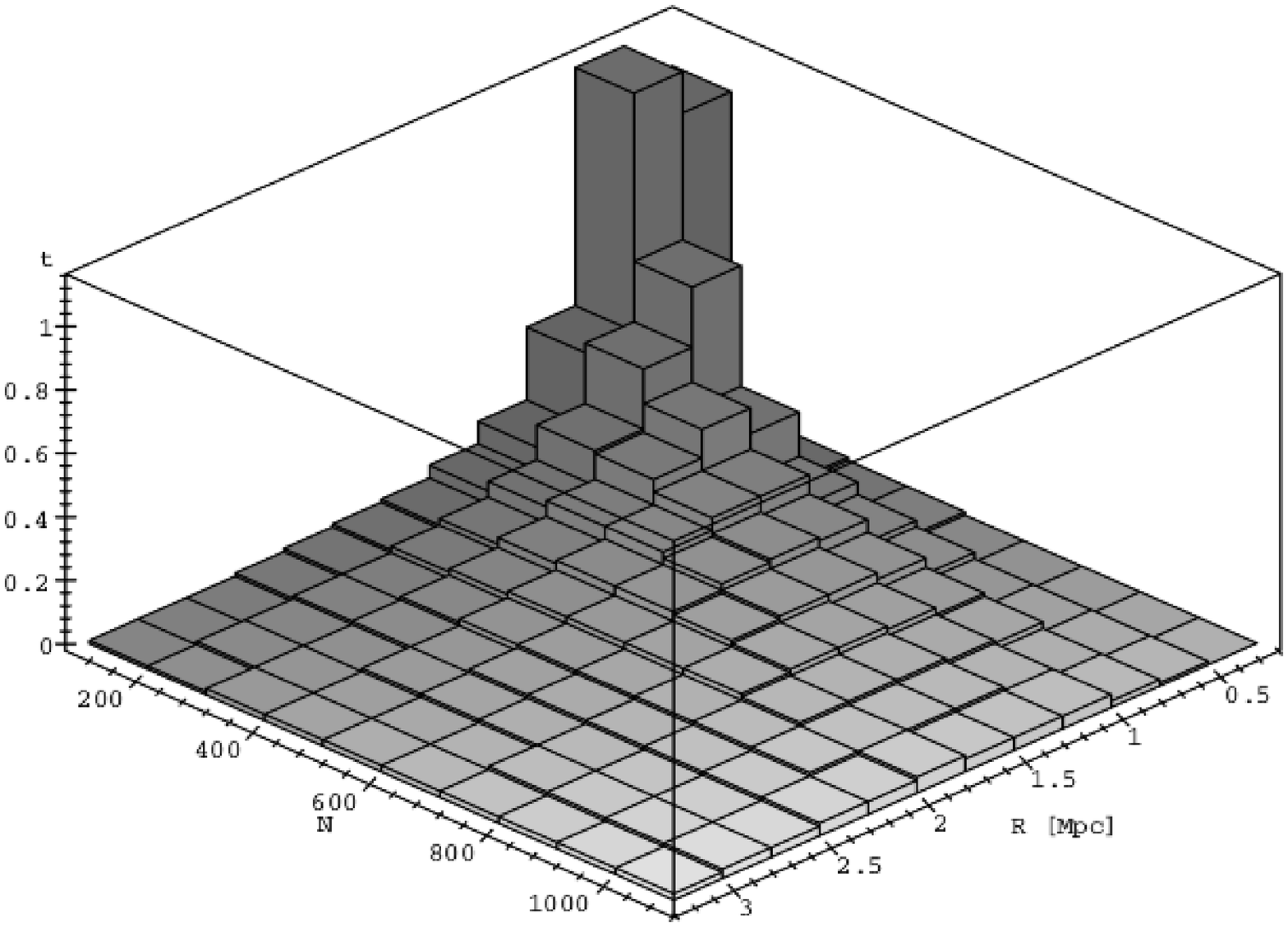}
\caption{Same as Fig.\ref{Abb8}, now for model \ref{simIII}.}
\label{Abb28}
\end{figure}

At smaller galaxy numbers the percentage of those galaxies that have been
involved in merger processes increases up to sixty per cent (Fig.~\ref{Abb28}),
cf. also Fig.~\ref{Abb8}. From a galaxy number of about four hundred onwards
there are practically no merger events to be noted.

\begin{figure}
{\includegraphics[angle=270,scale=0.56]{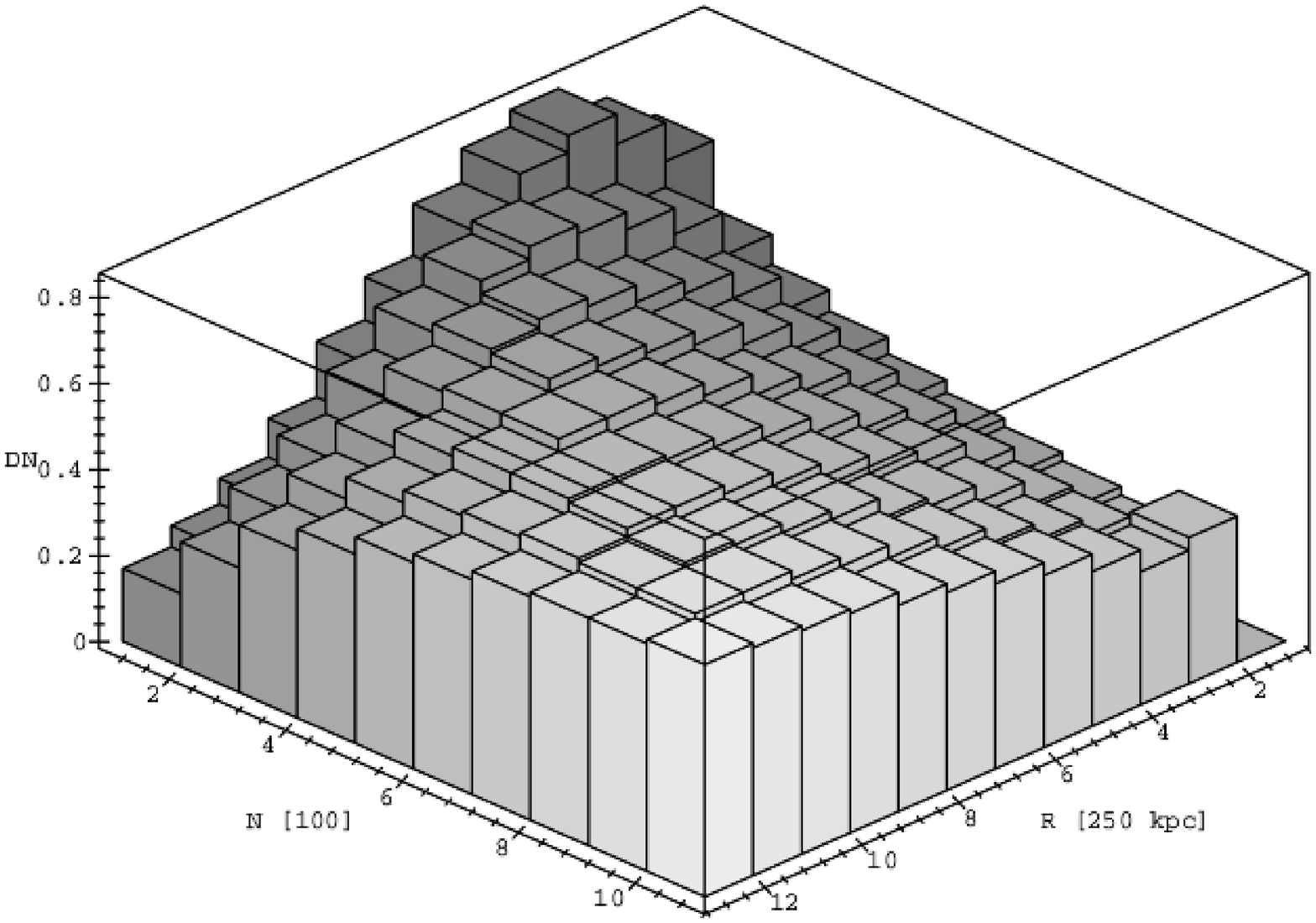}}
\caption{Same as Fig.\ref{Abb7}, now for model \ref{simIII}.}
\label{Abb17}
\end{figure}

\section{Discussion}

There are two processes with opposite effects which rule the evolution of the 
model: On the one hand the number of encounters decreases with increasing 
cluster radius at fixed galaxy number (cf. Fig.~\ref{Abb9}). On the other hand
the virial velocity increases with increasing total cluster mass by increasing 
richness at fixed cluster radius. Although the number of
encounters rises with velocity, the equations of energy transfer 
(Eqs.~\ref{spitzer1} and \ref{spitzer2}) depend on $v^{-2}$, and the kinetic
energy term of the relative motion depends on $v^2$. So high energy transfers are
only effective for mergers at low velocities. Taking both effects together, only
compact groups evolve noticeably in contrast to clusters. The number of 
encounters $g$ of a single galaxy for the total development time can be 
estimated to

\begin{equation}\label{simple1}
g =  \frac{N-1}{4\pi R^3/3}\pi B^2v T
\end{equation}
if richness does not change dramatically. The merger probability can be 
estimated in the same way: let $b_m$ be the impact parameter so that the merger
criterion is fulfilled within its enclosed area:

\begin{equation}
b_m=\sqrt{\frac{32}{3}}\frac{\sqrt{Gm_0r_0}}{v}.
\end{equation}
Then the fraction of merger events (caused by the stochastic choice of impact 
parameters) is $b_m^2/B^2$. This fraction multiplied by the number of
encounters then reveals the number of merger events $V$ of one galaxy 
(e.g. the BCG):

\begin{equation}\label{simple3}
V = 8 f^{-1}r_0T\sqrt{Gm_0}\cdot\sqrt{\frac{N}{R^5}}
\end{equation}

So the number $t$ of mergers depends strongly on the cluster radius and 
decreases with increasing $R$ (cf. Fig.~\ref{Abb7}). On the other hand $t$ 
increases proportional to the square root of the initial number of galaxies 
(assumed that it remains nearly constant, e.g. Fig.~\ref{Abb7} at $R=0.75$ Mpc 
and higher). The boost of constructive (remnant forming) or destructive 
encounters can thus be explained, but not the merger boost at low galaxy 
numbers. Furthermore it has to be taken into account whether the remnant 
remains bound. That leads to 

\begin{equation}
\frac{3\pi\beta}{8r_0f^2}\cdot\frac{R}{N} >1.
\end{equation}

The remnant has a chance to survive only at high cluster radii and small 
galaxy numbers. Together with the merger criterion this leads to the occurring
maximum development at small cluster radii and small numbers of galaxies 
(Figs.~\ref{Abb3}, \ref{Abb6}, \ref{Abb8}): Small radii are necessary to form
remnants, and their binding energies are negative only at small galaxy numbers.

These estimates expressed in simple formulae (Eqs.~\ref{simple1}-\ref{simple3})
are possible because our simulation show that there is no agravating change of
the initial mass distribution.

The simulation computations of Sect.~\ref{simIII} show a merger behaviour of
all galaxies and especially of those forming the BCG comparable to 
Sect.~\ref{simI}. A higher galaxy mass has a positive effect on the number of 
mergers because the mean energy transfer rises with the third power of the
characteristic mass, whereas the kinetic energy of the relative motion still 
rises linearly. On the other side a higher cluster mass in Sect.~\ref{simIII}
in contrast to Sect.~\ref{simI} causes the virial velocity to increase, 
suppressing energy transfer such as in Sect.~\ref{simII}.

Interpreting simulation \ref{simII} to represent cluster states further evolved 
in time compared with simulation \ref{simI} as the mass of the intergalactic 
medium increases due to tidal stripping (\cite{merritt84}), a picture of cluster evolution is 
formed: In the early phases of cluster evolution merger processes dominate the 
development of mass spectrum and BCG, then paling into insignificance as time goes by
(cf. \cite{albinger}). Comparing both simulations it should be noted
that the cluster mass is the same in simulation \ref{simII} at $N=100$ and 
simulation \ref{simI} at $N=1000$. An increase of the mass fraction of 
intergalactic matter leads to a drastic decrease in the merger rate at equal
galaxy mass distributions, because the mean kinetic energy controlled by the
virial theorem (Eq.~\ref{glvII}) outgrows the expected energy
transfers in Eq.~\ref{interpol}, so that mergers become improbable. The amount of 
galaxy binding energy is not influenced by a variation of $k$, and so a (less 
steep) decrease in galaxy destructions is expected on the basis of 
Eq.~\ref{interpol} as shown by the simulations.

A closer consideration of Fig.~\ref{Abb3} together with Fig.~\ref{Abb6}, 
dividing the mean increase of BCG mass by the mean number of mergers leading to
the BCG formation, shows that the brightest galaxy must have been built by
mergers of galaxies of uniform mass $M_0$ independent of general conditions
of the cluster. This is in accordance with the observation results of 
\cite{hoessel}.

Figure \ref{Abb8} gives information about the number of constructive mergers per
final galaxy number, i.e. the fraction of galaxies that merged with other 
galaxies during the simulation. This percentage is equivalent to the averaged 
number of mergers a galaxy has undergone. This fraction rises going from high to
small cluster radii and amounts to between some and about hundred per cent. 
Toomre (\cite{shu}) speculated that perhaps all elliptical cluster galaxies 
have been built by mergers. From the fraction of galaxies that are presently 
involved in gravitational interactions and the probable life time of this 
passing phenomena he calculated that merging of spiral galaxies happens for a 
fraction of about ten per cent of all cluster galaxies, in good agreement with 
the results of this work. The fact that especially rich compact clusters which 
are assumed to be dynamical older systems than spiral rich ones have a 
significantly larger fraction of elliptical galaxies (up to 35 per cent in the 
Coma cluster (cf. \cite{oemler})) may have multiple reasons. First a
comparison of the simulations in Sects.~\ref{simI} and \ref{simII} shows that
mergers grind to a halt at rising fraction of intergalactic matter, as the 
fraction of prospective interacting galaxies decreases at constant total cluster
mass due to mergers and destructions. Hence the merger rate is higher in a 
cluster of earlier state of development (corresponding to irregular spiral rich
clusters) than in the final state of compact clusters. So the fraction of 
elliptical galaxies in Toomre's estimation has to be corrected upwards. 
Secondly elliptical galaxies may be a result of disturbances caused by nearby
passages without merging that would lead to its rise to the presently observed
value.

The percentage of galaxies destroyed in encounters (Fig.~\ref{Abb10}) yields an
estimate of the dark matter fraction in clusters provided by destroyed galaxies.
This fraction amounts up to 60 per cent corresponding to a ratio of cluster mass
and total galaxy mass of $k=2.5$. 

Further simulations with the presented model were carried out using other sets
of parameters without leading to different results than discussed in this work. 
Particularly an increase in galaxy numbers beyond a value of 1000 does not 
change anything, because the evolution of any model with scales comparable to 
clusters takes an imperceptible course at high galaxy numbers.

\section{Conclusions}

There is a common result from all simulations of Sect.~\ref{results}: the 
cluster evolution is tightly related to the evolution of the brightest galaxy of
this cluster. Only at small galaxy numbers and simultaneously at small cluster 
radii the dynamical development due to the model described in Sect.\ref{model} 
plays a noticeable role (cf. e.g. Fig.~\ref{Abb6}). At the same time increasing 
merger action is a concomitant of increasing destruction of galaxies. So, in the
parameter range of rapid evolution, the number of galaxies depleted such that 
the evolved small cluster looks like a compact group rather than an Abell 
cluster. Nevertheless merging is always the dominant process (at least for the 
most massive galaxies). For clusters of presently observed richness and their 
most luminous member (being the comparative value to our simulation) the 
dynamical development of the mass spectrum and of correlated quantities as the 
BCG mass causes only little changes with respect to the initial conditions. 

Considering Fig.~\ref{Abb2} and Fig.~\ref{Abb4} the steadiness of BCG masses and
their ensemble dispersion attract attention (irrespective of compact groups). 
This approximate independence of masses on cluster radius is observed in 
reality, and there are two causes that balance each other: At the beginning of 
the simulation the BCG masses are small and rise with increasing richness. In 
the course of evolution mergers take place preferentially at low galaxy numbers 
resulting in a relatively constant BCG mass level after the development in 
dependence of the parameters $R$ and $N$. Taking into account an additional 
correlation between cluster radius and richness (e.g. \cite{oemler}) the 
BCG mass steadiness would turn out almost clearer, just as their dispersion 
would be narrower than stated in this work because of the restriction in
parameter space. 

Why does the BCG mass scatter remain significantly narrow even at the end of 
cluster development? First of all the number of mergers is relatively small
for the mentioned reasons (between one and six merger events depending on the
parameter set), and secondly the identity of a galaxy to be BCG is fixed only 
when the cluster development is complete. So, not the mass dispersion of a 
distinct galaxy fixed before the starting development is investigated, but the 
mass dispersion of a galaxy that is a posteriori identified to be the most 
massive one. Therefore the BCG mass dispersion turns out to be smaller than in 
the case of any individual galaxy chosen at the beginning of the simulation and
following its evolution.

\acknowledgements 
I thank Prof.~Dr.~R.~Wielen and Prof.~Dr.~B.~Fuchs for 
	stimulating discussions and Dr.~S.~Frink for help in the preperation of this
	work for publication.

\end{document}